\documentclass{article}
\usepackage{hiph-art}
\usepackage{graphicx}
\volnumber{00} \issuenumber{0} \edyear{2006}                             
\frompage{000} \topage{000}                                              
\recrevdate{15 November 2005}                                            
\def\rightmark{Symmetry Energy in the EOS of Asymmetric Nuclear Matter}
\def\leftmark{Yennello, Shetty, Souliotis}
 
\title{Symmetry Energy in the Equation of State of Asymmetric Nuclear Matter \\}
\authors{ 
{S.J. Yennello$^1$, D.V. Shetty$^{1}$ and G.A. Souliotis$^{1}$ %
\index{Yennello, S.J.} 
\index{Shetty, D.V.} 
\index{Souliotis, G.A.} 
}\\[2.812mm]
{\normalsize
\hspace*{-8pt}$^1$ Cyclotron Institute, Texas A$\&$M University \\
[0.2ex] 
\\ 
College Station, TX 77843, USA
}}
 
\abstract{The symmetry energy is an important quantity in the equation of 
state of isospin asymmetric nuclear matter. This currently unknown quantity is 
key to understanding the structure of systems as diverse as the neutron-rich nuclei 
and neutron stars. At TAMU, we have carried out studies, aimed at understanding the 
symmetry energy, in a variety of reactions such as, the multifragmentation of 
$^{40}$Ar, $^{40}$Ca + $^{58}$Fe, $^{58}$Ni and $^{58}$Ni, $^{58}$Fe + $^{58}$Ni, $^{58}$Fe 
reactions at 25 - 53 AMeV, and deep-inelastic reactions of $^{86}$Kr + $^{124,112}$Sn, $^{64,58}$Ni (25 AMeV), $^{64}$Ni + $^{64,58}$Ni,
$^{112,124}$Sn, $^{232}$Th, $^{208}$Pb (25 AMeV) and $^{136}$Xe + $^{64,58}$Ni, $^{112,124}$Sn,
$^{232}$Th, $^{197}$Au (20 AMeV). Here we present an overview of some of the 
results obtained from these studies. The results are analyzed within the framework of 
statistical and dynamical models, and have important implications for future experiments 
using beams of neutron-rich nuclei.}

 
\makeindex   
\begin{document}
 
\maketitle

\section{Introduction}
The equation of state of isospin asymmetric nuclear matter is a fundamental 
quantity that determines the properties of systems as small and light as an
atomic nucleus, and as large and heavy as a neutron star. The key unknown in
the EOS of asymmetric nuclear matter is the symmetry energy. Recently the possibility 
of extracting information on the symmetry energy and the isospin (neutron-to-proton ratio) 
of the fragments in a multifragmentation reaction has gained tremendeous importance
\cite{SHE05,FEV05}. Such information is of importance for understanding key 
problems in astrophysics\cite{DAN02}, and various aspects of nuclear physics such as the structure 
of exotic nuclei (the binding energy and rms radii) \cite{BRO00,HORO01} and the dynamics of 
heavy ion collisions \cite{BAL01}.
\par
Traditionally, the symmetry energy of nuclei has been extracted 
by fitting the binding energy of the ground state with various versions of the liquid drop 
mass formula. The properties of nuclear matter are then determined by 
theoretically extrapolating the nuclear models designed to study the structure of real nuclei. 
However, real nuclei are cold, nearly symmetric ($N \approx Z$) and found at equilibrium density. It is
not known how the symmetry energy behaves at temperatures, isospin (neutron-to-proton ratio) and 
densities away from the normal nuclear matter. Theoretical many-body calculations \cite{DIE03} 
and those from the empirical liquid drop mass formula \cite{MYE66} predict symmetry 
energy near normal nuclear density ($\approx$ 0.17 fm$^{-3}$) and temperature ($T$ $\approx$ 0 MeV), 
to be around 28 - 32 MeV. 
\par
In a multifragmentation reaction, an excited nucleus expands to a sub-nuclear density and disintegrates 
into various light and heavy fragments. The fragments are highly excited and neutron-rich ; their 
yields depend on the available free energy, which in turn depends on the strength of the symmetry 
energy and the extent to which the fragments expand. By studying the isotopic yield distribution 
of these fragments, one can extract important information about the symmetry energy and the properties 
of the fragments at densities, excitation energies and isospin away from those of ground state nuclei. 
Here we present some of the results obtained from various measurements carried out at the Cyclotron
Institute of Texas A$\&$M University (TAMU). We present these results in the framework of both, the 
statistical and the dynamical multifragmentation models. 

\begin{figure}
\includegraphics[width=0.52\textwidth]{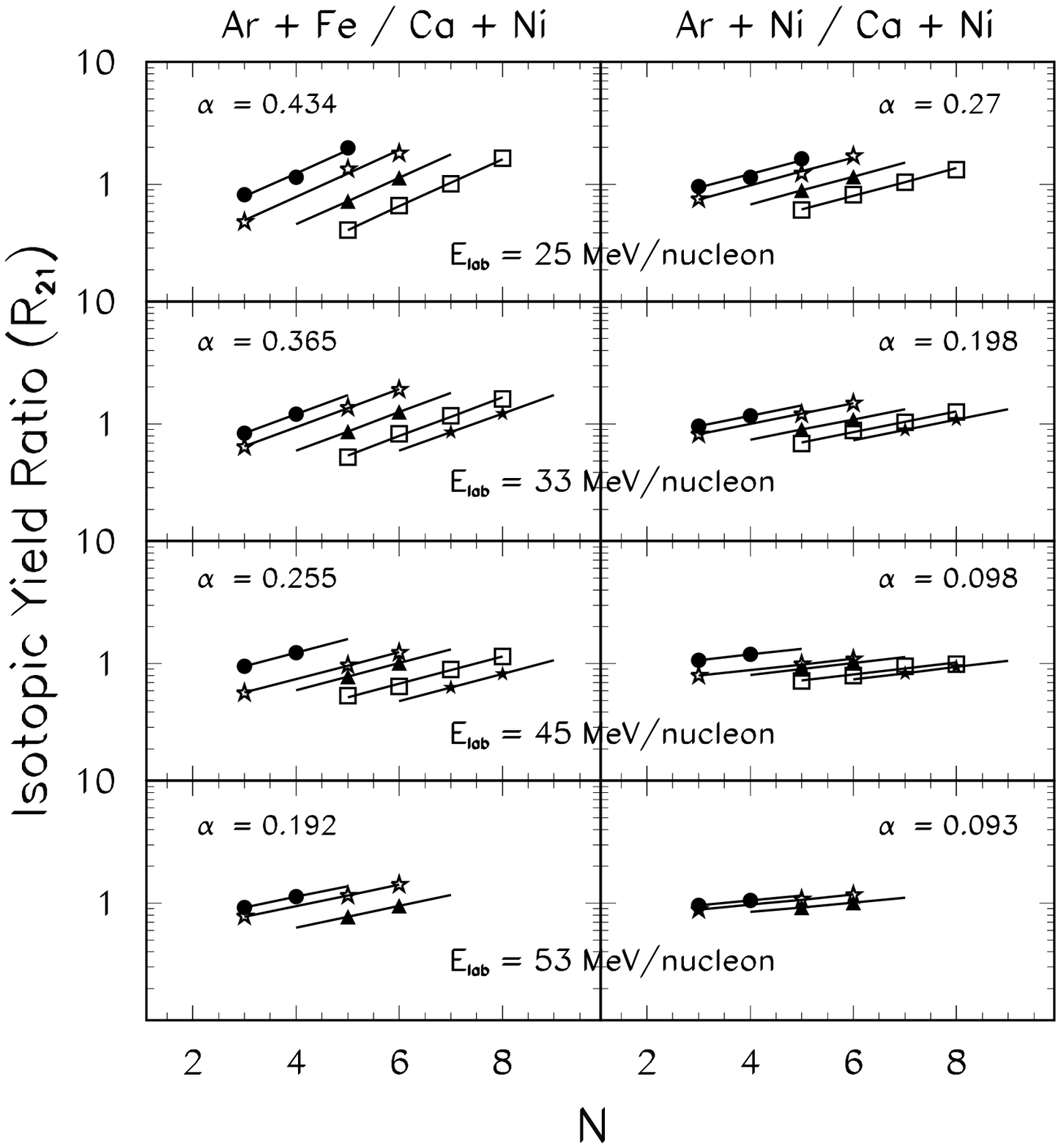}
\includegraphics[width=0.45\textwidth, height=0.40\textheight]{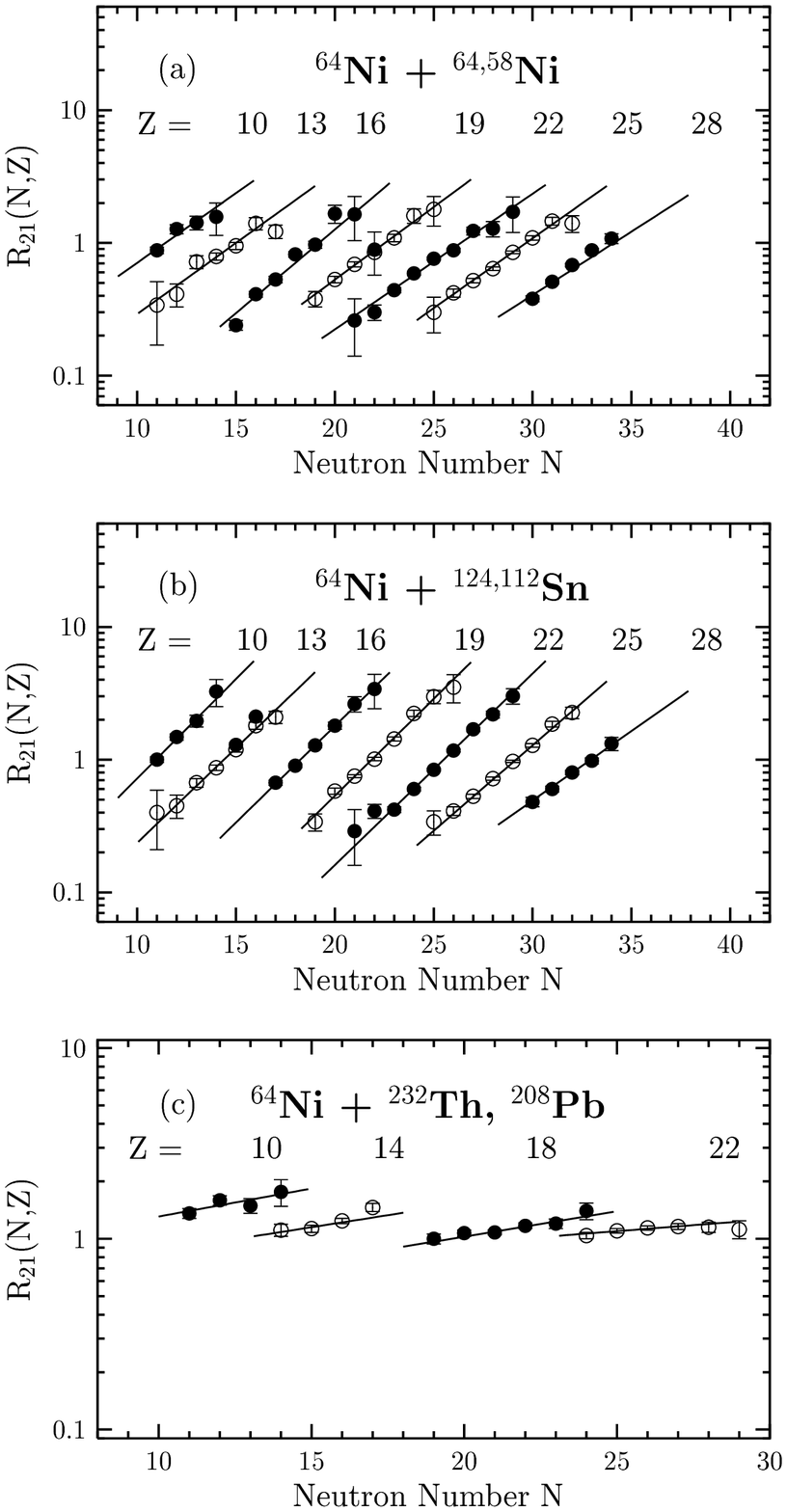}
\caption{Isotopic yield ratios for $^{40}$Ar, $^{40}$Ca + $^{58}$Fe, $^{58}$Ni reactions at 25 - 53
MeV/nucleon, and deep-inelastic reactions of $^{64}$Ni + $^{64,58}$Ni, $^{112,124}$Sn, $^{232}$Th,
$^{208}$Pb (25 MeV/nucleon)}
\end{figure} 

\section{Isoscaling and Symmetry Energy}
In a multifragmentation reaction, the ratio of isotope yields in two different reactions, 1
and 2, $R_{21}(N,Z) = Y_{2}(N,Z)/Y_{1}(N,Z)$, has been shown to obey an exponential dependence on the
neutron number ($N$) and the proton number ($Z$) of the isotopes, an observation known as isoscaling
\cite{BOT02,TSAN01,TSANG01}. The dependence is characterized by a simple relation,

\begin{equation}
              R_{21}(N,Z) = Y_{2}(N,Z)/Y_{1}(N,Z) = C exp({\alpha N + \beta Z})
\end{equation}

where, $Y_{2}$ and $Y_{1}$ are the yields from the neutron-rich and neutron-deficient systems,
respectively. $C$ is an overall normalization factor, and $\alpha$ and $\beta$ are the parameters
characterizing the isoscaling behavior. 
\par
Theoretically, isoscaling has been predicted by both, statistical \cite{TSAN01} as well as
dynamical \cite{ONO03} multifragmentation models.
In these models, the difference in the chemical potential of systems with different $N/Z$ is directly 
related to the scaling parameter $\alpha$. The scaling parameter $\alpha$ is 
proportional to the symmetry energy through the relation,

\begin{equation}
   \alpha = \frac{4 C_{sym}}{T} \bigg (\frac{Z_{1}^{2}}{A_{1}^{2}} - \frac{Z_{2}^{2}}{A_{2}^{2}} \bigg )
\end{equation}

In the above equation, $Z_{1}$, $A_{1}$ and $Z_{2}$, $A_{2}$ are the charge and the mass numbers of the 
fragmenting systems, $T$ is the temperature of the system and $C_{sym}$, the symmetry energy.
The parameter $\alpha$, has been shown to be independent of the complex nature of the secondary
de-excitation of the primary fragments, and is thus a robust observable for studying the symmetry 
energy \cite{TSAN01}. While it is well established that many versions of statistical models show very 
little or no difference between the $\alpha$ values  for the primary and the secondary fragments, the same
may not be true for the $\alpha$ values obtained from dynamical models. The origin of this
discrepancy between the two approaches is currently being debated and not fully understood. It 
will be shown from the present study that the difference between the primary and the secondary $\alpha$'s 
in statistical model is very small. We will presume the difference to be insignificant in the dynamical 
model framework.   
\par
Fig. 1 shows the experimentally determined isotopic yield ratio as a function of neutron number $N$, for
some of the reactions studied using beams from the K500 Cyclotron at TAMU. The figure on the left shows 
the ratios for the ($^{40}$Ar + $^{58}$Fe)/($^{40}$Ca + $^{58}$Ni) and ($^{40}$Ar +
$^{58}$Ni)/($^{40}$Ca + $^{58}$Ni) pairs 
of reactions. The one on the right is for the deep-inelastic reactions of $^{64}$Ni + Ni, Sn, Th, Pb at
20 AMeV. One observes that the ratios for various elements in a given reaction pair lie along a 
straight line in the logarithmic plot and align with the neighboring elements in accordance with the relation 
given in equation 1. This feature is observed for all the beam energies and the pairs of reactions studied. The 
alignment of the data points varies with beam energies as well as the pairs of reaction. The ratio for each 
elements ($Z$) were simultaneously fitted using an exponential relation (shown by the solid lines) to obtain the 
slope parameter $\alpha$. In the following sections, we use these experimentally determined $\alpha$'s to study the 
symmetry energy using the statistical and dynamical model interpretation of the multifragmentation reaction.
 
\section{Symmetry Energy from a Statistical Model Approach}  
The Statistical Multifragmentation Model (SMM) \cite{BON95,BOT95} is the most widely used model for describing 
multfragmentation 
reactions. It is based on the assumption of statistical equilibrium at a low density freeze-out stage.
All breakup channels composed of nucleons and excited fragments are taken into account and considered 
as partitions.  During each partition the conservation of mass, charge, energy and angular momentum 
is taken into account, and the partitions are sampled uniformly in the phase space according to their 
statistical weights using Monte Carlo sampling. The Coulomb interaction between the fragments is 
treated in the Wigner-Seitz approximation. Light fragments with mass number $A$ $\leq$ 4 are considered 
as elementary particles with only translational degrees of freedom (``nuclear gas"). Fragments with $A$ $>$ 4 
are treated as heated nuclear liquid drops, and their individual free energies $F_{A,Z}$ are parametrized as 
a sum of the volume, surface, Coulomb and symmetry energy,

\begin{equation}
            F_{A,Z} = F^{V}_{A,Z} + F^{S}_{A,Z} + E^{C}_{A,Z} + E^{sym}_{A,Z}
\end{equation}

where $F^{V}_{A,Z} = (-W_{o} - T^{2}/\epsilon_{o})A,$ with parameter $\epsilon_{o}$ related to the level
density and $W_{o}$ = 16 MeV being the binding energy of infinite nuclear matter. $F^{S}_{A,Z} = B_{o}A^{2/3}[(T^{2}_{c} -
T^{2})/(T^{2}_{c} + T^{2})]^{5/4}$, with $B_{o}$ = 18 MeV being the surface co-efficient and $T_{c}$ = 18 MeV being the 
critical temperature of infinite nuclear matter. $E^{C}_{A,Z} = c Z^{2}/A^{1/3}$, where $c = (3/5)(e^{2}/r_{o})[1 - (\rho/\rho_{o})^{1/3}]$ 
is the Coulomb parameter obtained in the Wigner-Seitz approximation with charge unit $e$, and $r_{o}$ = 1.17 fm. 
$E^{sym}_{A,Z} = C_{sym}(A - 2Z)^{2}/A$, where $C_{sym}$ = 25 MeV is the symmetry energy co-efficient. These 
parameters are adopted from the Bethe-Weizsacker mass formula and correspond to the assumption of isolated 
fragments with normal density in the freeze-out 
configuration. The value of the symmetry energy co-efficient $C_{sym}$ is taken from the fit to the binding energies of 
isolated cold nuclei in their ground states. In a multifragmentation process the  primary fragments are not only 
excited but also expanded. 
\par
Figure 2 shows a comparison between the SMM calculated and the experimentally 
observed values of $\alpha$. The left side of the figure corresponds to the  ($^{40}$Ar +
$^{58}$Ni)/($^{40}$Ca + $^{58}$Ni) and the ($^{40}$Ar + $^{58}$Fe)/($^{40}$Ca + $^{58}$Ni) 
pairs of reactions. The one on the right corresponds to ($^{58}$Fe + $^{58}$Fe)/($^{58}$Ni + $^{58}$Ni)
and ($^{58}$Fe + $^{58}$Ni)/($^{58}$Ni + $^{58}$Ni) pairs of reactions. The dotted lines in the $\alpha$ versus 
excitation energy plot corresponds to $\alpha$ calculated from the primary fragment
distribution and the solid lines to those calculated from the secondary fragment distribution. The
symbols correspond to the experimentally determined $\alpha$'s.  The figure on the left clearly
shows that the experimentally 
determined $\alpha$'s are significantly lower than the calculated values using the standard 
value of the symmetry energy, $C_{sym}$ = 25 MeV. To explain the observed dependence of the isoscaling 
parameter  $\alpha$ on
excitation energy, the $C_{sym}$ of the hot primary fragment in the SMM calculation was varied in the range 25 - 15 MeV. 
As shown in the center and the bottom panel of the left figure, the isoscaling parameter decreases slowly
with decreasing symmetry energy. The experimentally determined $\alpha$ can be reproduced for both 
pairs of systems at all excitation energies using a symmetry energy value of $C_{sym}$ = 15 MeV.
This value of the symmetry energy is significantly lower than the value of $C_{sym}$ = 25 MeV often
used for the isolated cold nuclei in their ground states. On the right side of the figure, we show
the comparisons for the ($^{58}$Fe + $^{58}$Fe)/($^{58}$Ni + $^{58}$Ni) and  ($^{58}$Fe +
$^{58}$Ni)/($^{58}$Ni + $^{58}$Ni) pairs
of reactions. Once again a lower value of symmetry energy $C_{sym}$ is required to explain the experimental
data. Furthermore, one also observes a small dependence of the symmetry energy with increasing
excitation energy.
 
\begin{figure}
\includegraphics[width=0.52\textwidth,height=0.34\textheight]{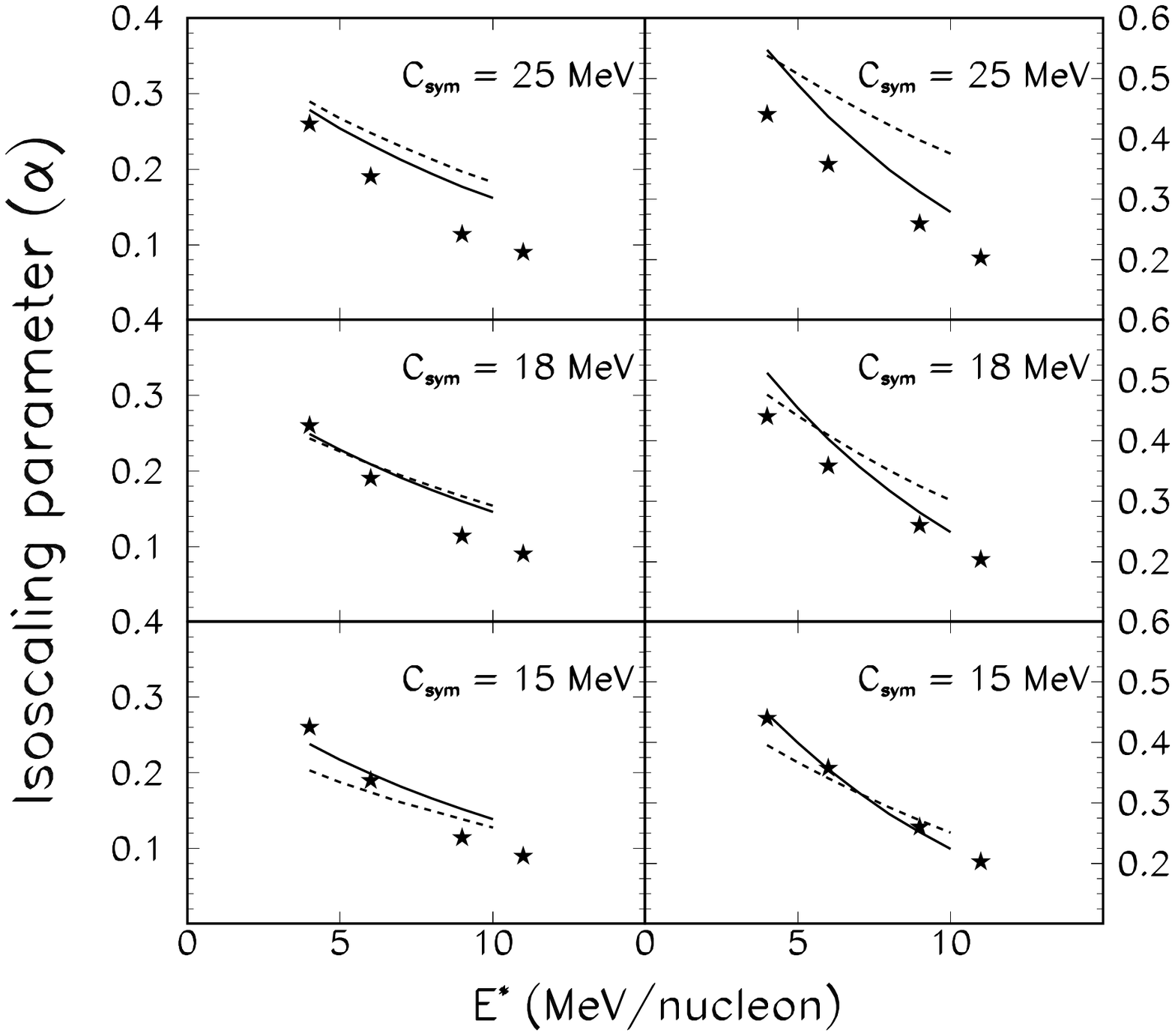}
\includegraphics[width=0.52\textwidth]{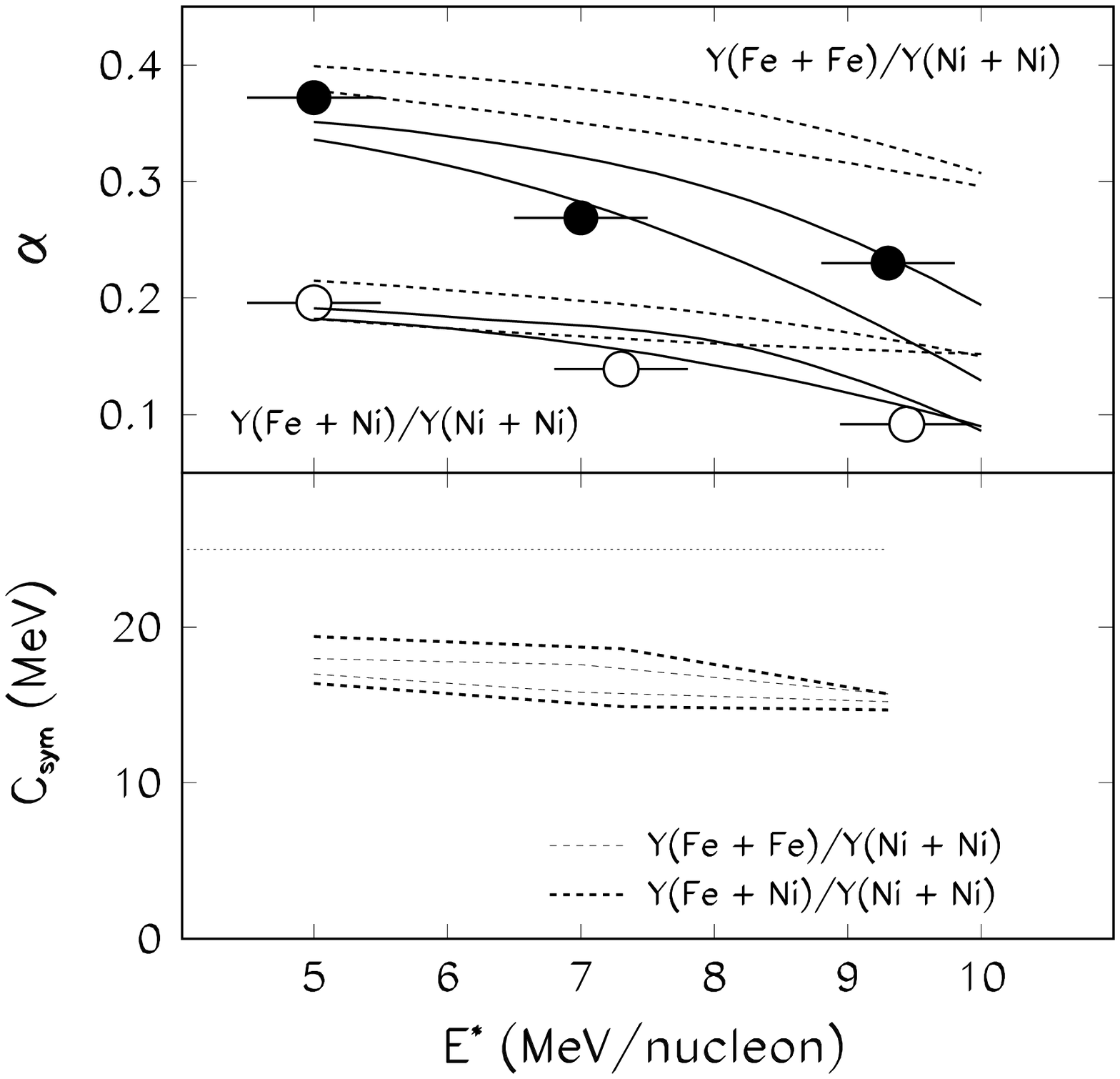}
\caption{Statistical model comparison of the isoscaling parameter with the experimentally
determined values as a function of the excitation energy.} 
\end{figure} 
\begin{figure}
\begin{center}
\includegraphics[width=0.52\textwidth]{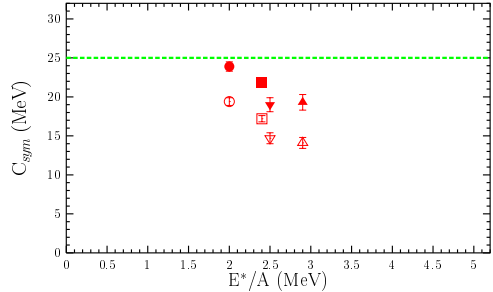}
\end{center} 
\caption{Symmetry energy as a function of the excitation energy for $^{86}$Kr + $^{124,112}$Sn
(circles), $^{64,58}$Ni (squares) (25 AMeV), $^{64}$Ni + Ni, Sn, Th, 
Pb (triangles) (25 AMeV) and $^{136}$Xe + Ni, Sn, Th, Au (inverted triangles) (20 AMeV) reactions.}
\end{figure} 
  
\par
A similar behavior of the $C_{sym}$ is also observed from the deep-inelastic reactions studies of 
$^{86}$Kr + $^{124,112}$Sn, $^{64,58}$Ni (25 AMeV), $^{64}$Ni + Ni, Sn, Th, Pb (25 AMeV) 
and $^{136}$Xe + Ni, Sn, Th, Au (20 AMeV) \cite{SOU05}. 
Fig. 3 shows the values of the $C_{sym}$ obtained from these reactions. The values were obtained
using equation 2 with two different assumptions for temperature determination ; the Fermi gas 
temperature (closed symbols) and the expanding mononucleus temperatures(open
symbols). The symmetry values for both set of temperatures show decreasing trend with increasing
excitation energy. 
\par
The above comparison of the experimentally observed isoscaling parameter with the statistical
multifragmentation model therefore shows that a significantly lower value of the symmetry energy is required
to explain the isotopic composition of the fragments produced in a fragmentation reaction. This
indicates that the properties of nuclei at high excitation energy, isospin and reduced density are
very sensitive to the symmetry energy. Similar hot and neutron-rich nuclei are routinely produced in the
interior of a collapsing star and subsequent supernova explosion, where a slight decrease in the symmetry
energy can significantly alter the elemental abundance and the synthesis of heavy elements
\cite{BOT04}. The
present observations can provide important inputs for the understanding of the nuclear composition
of supernova matter.

\section{Symmetry Energy from a Dynamical Model Approach}
In the following, we analyze the above results in the dynamical model framework using the Anti-symmetrized
Molecular Dynamic (AMD)
calculation \cite{ONO03}. AMD is a microscopic model that simulates the time evolution of a nuclear collision, where the
colliding system is represented in terms of a fully antisymmetrized product of Gaussian wave packets.
During the evolution, the wave packet centroids move according to the deterministic equation of motion.
The followed state of the simulation branches stochastically and successively into a huge number of
reaction channels. The interactions are parameterized in terms of an effective force acting between
nucleons and the nucleon-nucleon collision cross-sections.  
The beauty of the dynamical models is that it allows one to understand the
functional form of the density dependence of the symmetry energy at a very fundamental level i.e., 
from the basic nucleon-nucleon interactions.   Theoretical 
studies \cite{DIE03} based on microscopic many-body calculations 
and phenomenological approaches predict various forms of the density dependence of the 
symmetry energy. In general, two different forms have been identified. One, where the 
symmetry energy increases monotonically with increasing density (`` stiff " dependence) and the other, 
where the symmetry energy increases initially up to normal nuclear density and then decreases at 
higher densities (`` soft " dependence).
\par
Determining the exact form of the density dependence of the symmetry energy is important for studying 
the structure of neutron-rich nuclei \cite{BRO00,HORO01}, and studies relevant to 
astrophysical problems, such as the structure of neutron stars and the dynamics of supernova 
collapse \cite{LAT01}. For example, a `` stiff " density dependence 
of the symmetry energy is predicted to lead to a large neutron skin thickness compared to a `` soft " 
dependence \cite{OYA98}. Similarly, a `` stiff " dependence of the symmetry 
energy can result in rapid cooling of a neutron star, and a larger neutron star radius, compared to a soft 
density dependence \cite{LAT94}.
\par
Recently, a linear relation between the 
isoscaling parameter $\alpha$, and the difference in the isospin asymmetry ($Z/A)^{2}$ of the fragments, 
with appreciably different slopes, was predicted for two different forms of the density dependence of the 
symmetry energy ; a `` stiff " dependence (obtained from Gogny-AS interaction) and a `` soft " dependence 
(obtained from Gogny interaction). 
\par
Fig. 4 (left) shows  a comparison between the experimentally observed $\alpha$ and those from the
AMD model calculations plotted as a function of the difference in the fragment 
asymmetry for the beam energy of 35 MeV/nucleon. The solid and the dotted lines are the AMD 
predictions using the `` soft " (Gogny) and the `` stiff " (Gogny-AS) density dependence of the 
symmetry energy, respectively.
The solid and the hollow symbols (squares, stars, triangles and circles) are the results of  the present 
study for the two different values of the fragment asymmetry, assuming Gogny and Gogny-AS 
interactions, respectively. Also shown in the figure are the scaling parameters (asterisks, crosses, diamond 
and inverted triangle) taken from various other works in the literature. It is
observed that the experimentally determined $\alpha$ parameter increase linearly with increasing 
difference in the asymmetry of the two systems as predicted by the AMD calculation. Also, the data points 
are in closer agreement with those predicted by the Gogny-AS interaction (dotted line) than those from the 
usual Gogny force (solid line). The slightly lower values of the symbols from the present measurements 
with respect to the Gogny-AS values (dotted line) could be due to the small secondary de-excitation
effect of the fragments not accounted for in this comparison. 
It has been shown \cite{TSANG01} that the experimentally determined $\alpha$ values can be lower 
by about 10 - 15 $\%$ for the systems and the energy studied here. Accounting for this effect results in 
a slight increase in the $\alpha$ values bringing them even closer to the dotted line. 
The observed agreement of the experimental data with the Gogny-AS type of interaction therefore appears to suggest 
a stiffer density dependence of the symmetry energy. However, as mentioned in section 2, the effect
of secondary de-excitation in the dynamical model calculations is currently under study \cite{ONO} and the
predicted sensitivity may be significantly diminished by the secondary decay.
\par
Recently, Chen {\it {et al.}} \cite{CHE05} also showed, using the isospin dependent Boltzmann-Uehling-Uhlenbeck 
(IBUU04) transport model calculation, that a stiff density dependence of the symmetry energy 
parameterized as E$_{sym}$ $\approx$ 31.6 ($\rho/\rho_{\circ})^{1.05}$ explains well the isospin 
diffusion data \cite{TSA04} from NSCL-MSU (National Superconducting Cyclotron Laboratory at Michigan 
State University).  Their  calculation was also based on a momentum-dependent Gogny effective 
interaction. However, the present measurements on isoscaling gives a slightly softer density dependence 
of the symmetry energy at higher densities than those obtained by Chen {\it {et al.}}
\par
The difference in stifness is clear from figure 4 (right), which shows the parameterization of various theoretical predictions of the 
density dependence of the nuclear symmetry energy in isospin asymmetric nuclear matter. The 
dot-dashed, dotted and the dashed curve corresponds to those from the momentum dependent 
Gogny interactions used by Chen {\it {et al.}} to explain the isospin diffusion data. These are given 
as, E$_{sym}$ $\approx$ 31.6 ($\rho/\rho_{\circ})^{\gamma}$, where, $\gamma$ = 1.6, 1.05 and 0.69, 
respectively. The solid curves and the solid points correspond to those from the Gogny and 
Gogny-AS interactions used to compare with the present isoscaling data. As shown by Chen {\it {et al.}}, 
the dependence parameterized by E$_{sym}$ $\approx$ 31.6 ($\rho/\rho_{\circ})^{1.05}$ (dotted curve)
explains the NSCL-MSU data on isospin diffusion quite well. On the other hand, the isoscaling data
from the present work can be explained well by the Gogny-AS interaction (solid points).

\begin{figure}
\includegraphics[width=0.52\textwidth]{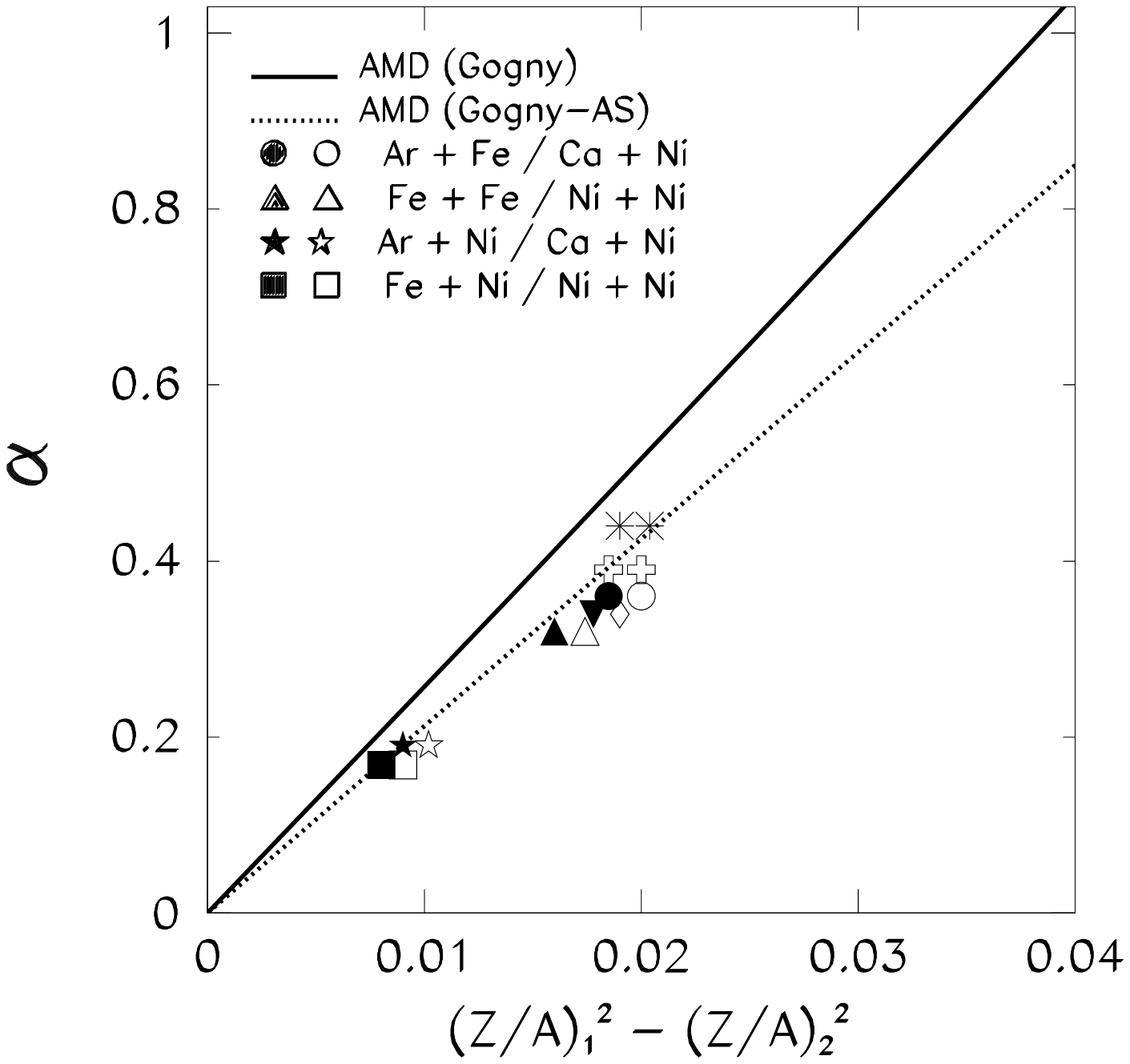}
\includegraphics[width=0.52\textwidth]{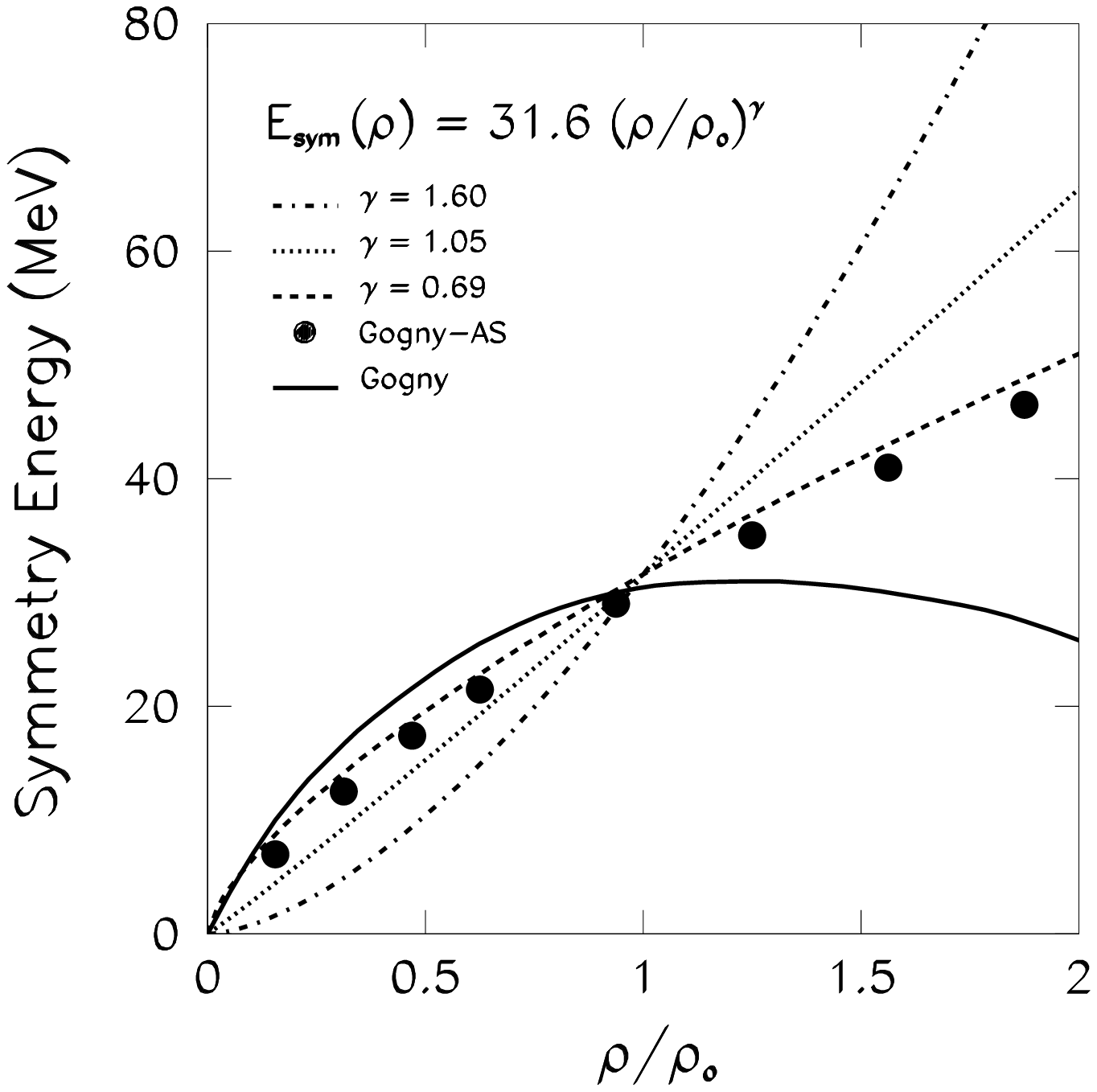}
\caption{(left) Scaling parameter as a function of the difference in the fragment asymmetry for 35
MeV/nucleon. (Right) Parameterization of the various forms of the density dependence of the nuclear
symmetry energy used in the analysis.}
\end{figure}

Both measurements yield similar results at low densities with significant difference at higher densities. 
It is interesting to note that by parameterizing the density dependence of the symmetry energy that 
explains the present isoscaling data, one gets, E$_{sym}$ $\approx$ 31.6 ($\rho/\rho_{\circ})^{\gamma}$, 
where $\gamma$ = 0.69. This form of the density dependence of the symmetry energy is consistent 
with the parameterization adopted by Heiselberg and Hjorth-Jensen in their studies on neutron stars \cite{HEI00}. 
By fitting earlier predictions of the variational calculations by Akmal {\it {et al.}} \cite{AKM98}, 
where the many-body and special relativistic corrections are progressively incorporated, Heiselberg 
and Hjorth-Jensen obtained a value of E$_{sym}$($\rho_{\circ}$) = 32 MeV and $\gamma$ = 0.6, 
similar to those obtained from the present measurements. The present form of the density dependence is 
also consistent with the findings of Khoa {\it {et al.}} \cite{KHO05}, where a comparison of the 
experimental cross-sections in a charge-exchange reaction with the Hartree-Fock calculation 
using the CDM3Y6 interaction \cite{KHO97}, reproduces well the empirical half-density point of 
the symmetry energy obtained from the present work (see fig. 2 of Ref. \cite{KHO05}).
\par
The observed difference in the form of the density dependence of the symmetry energy between the 
present measurement and those obtained by Chen  {\it {et al.}} is not surprising. Both measurements 
probe the low density part of the symmetry energy and are thus less sensitive to the high density 
region. But the important point to be noted is that both measurements clearly favor a stiff density 
dependence of the symmetry energy at higher densities, ruling out the very `` stiff '' 
(dot-dashed curve) and very `` soft '' (solid curve) predictions. These results can thus be used 
to constrain the form of the density dependence of the symmetry energy at supranormal densities 
relevant for the neutron star studies.
\par
In view of the findings from the present measurements and those of Chen {\it {et al.}}, we believe 
that the best estimate of the density dependence of the symmetry energy that can be presently 
extracted from heavy ion reaction studies is, E$_{sym}$ $\approx$ 31.6 ($\rho/\rho_{\circ})^{\gamma}$, 
where $\gamma$ = 0.6 - 1.05. Measurements at higher densities should be able to constrain the 
density dependence of the symmetry energy further. 

\section{Conclusions}
In conclusion, a number of studies have been carried out at TAMU to study the symmetry energy in
the equation of state of isospin asymmetric nuclear matter. 
The results were analyzed within the framework of statistical and dynamical model calculations.
It is observed that the
properties of nuclei at excitation energy, isospin and density away from the normal ground state
nuclei are significantly different and sensitive to the symmetry energy. 
The symmetry energy
required to explain the isoscaling parameter of the fragments produced in multifragmentation
reactions are significantly lower, and as small as 15 MeV.  
The dynamical model calculation of the isoscaling parameter shows that a stiffer form of the 
density dependence of the symmetry energy is preferred over a soft dependence. A dependence of 
the form E$_{sym}$ $\approx$ 31.6 ($\rho/\rho_{\circ})^{0.69}$ 
appears to agree better with the present data. Recently it has been shown that this form of
the density dependence of symmetry energy provides an accurate description of several collective
modes having different neutron-to-proton ratios. Among the predictions from this dependence are a
symmetric nuclear-matter incompressibility of $K$ = 230 MeV and a neutron skin thickness in
$^{208}$Pb of 0.21 fm. Further, this dependence leads to a neutron star mass of $M_{max}$ = 1.72
$M_{\odot}$ and a radius of $R$ = 12.66 km for a ``canonical" $M$ = 1.4 $M_{\odot}$ neutron star. These 
results have significant implications for nuclear astrophysics and future experiments probing the 
properties of nuclei using beams of neutron-rich nuclei.

\section*{Acknowledgment(s)}
This work was supported in parts by the Robert A. Welch Foundation (grant No. A-1266) and the 
Department of Energy (grant No. DE-FG03-93ER40773). We also thank A. Botvina for fruitful discussions 
on statistical multifragmentation model.

\vfill\eject
\end{document}